# Public Transport Networks under Random Failure and Directed Attack

Bertrand Berche[(1)], Christian von Ferber[(2,3)], Taras Holovatch[(1,2)], Yurij Holovatch[(4)]

[(1)]Statistical Physics Group, P2M Dpt, Institut Jean Lamour, Nancy Université, BP 70239, F-54506 Vandœuvre les Nancy, France
berche@lpm.u-nancy.fr
[(2)] Statistical Physics Group, Applied Mathematics Research Centre, Coventry University, Coventry CV1 5FB, UK
holovatch@gmail.com
[(3)]Physikalisches Institut, Universität Freiburg, D-79104 Freiburg, Germany
c.vonferber@coventry.ac.uk
[(4)]Institute for Condensed Matter Physics, National Acad. Sci. of Ukraine, UA-79011 Lviv, Ukraine & Institut für Theoretische Physik, Universität Linz, A-4040, Linz, Austria
hol@icmp.lviv.ua

**Abstract.** The behaviour of complex networks under failure or attack depends strongly on the specific scenario. Of special interest are scale-free networks, which are usually seen as robust under random failure but appear to be especially vulnerable to targeted attacks. In a recent study of public transport networks of 14 major cities of the world we have shown that these systems when represented by appropriate graphs may exhibit scale-free behaviour. In this paper we briefly review some of the recent results about the effects that defunct or removed nodes have on the properties of public transport networks. Simulating different directed attack strategies, we derive vulnerability criteria that result in minimal strategies with high impact on these systems.

## 1. Introduction

Since the last decade we witness how the ideas and methods of the graph theory merge with those of statistical physics and give rise to complex network science [1]. The emergence of complex network theory is accompanied by a very fruitful implementation in different branches of natural and social sciences and even in humanities. Realization of the fact that many natural and man-made structures have a network topology and in many instances such networks have much in common, allows for an understanding of different phenomena that occur on such networks and for their quantitative description. The specific type of networks we address in this paper are public transport networks [2-14], that provide an instance of the more general class of transportation networks. Whereas a network topology is not apparent for many structures and can be recovered only in the course of a thorough analysis of their subtle features, in the case of public transportation the network structure is obvious and it is even fixed in the commonly used term „public transport network" (PTN). However, as we will see below, there are many ways to represent a PTN in the form of a graph [2,4,5,10,15], some of them will be used in our study. One can discriminate between two main directions of research in the field of complex networks: on the one hand, one is interested in the structural properties of complex networks, on the other hand, one analyzes different phenomena that occur on networks. Our recent analysis of PTN [10-14] shares goals of both of these directions: on the one hand, we analyze topological properties of PTN of 14 major cities of the world [10,11], on the other hand, we address particular processes that may and do occur on PTN [12-13]. Concerning the latter case we analysed effects that defunct or removed nodes have on the properties of PTN. Simulating different directed attack strategies, we derived vulnerability criteria that result in minimal strategies with high impact on these systems. In this paper we briefly review some of the recent results about behaviour of PTN under random failure or directed attack.

## 2. PTN representation

The question how the characteristics of a complex network change when some of its constituents are removed has many important practical implementations. Below, we will call such removal an *attack*. The notion of attack vulnerability of a network originates from studies of computer networks and denotes the decrease of network performance that is caused by removal of its nodes and/or links. In the case of a PTN, knowledge about its vulnerability allows both to take measures to protect its most vulnerable components as well as to possibly increase its resilience by planning its further development and evolution.

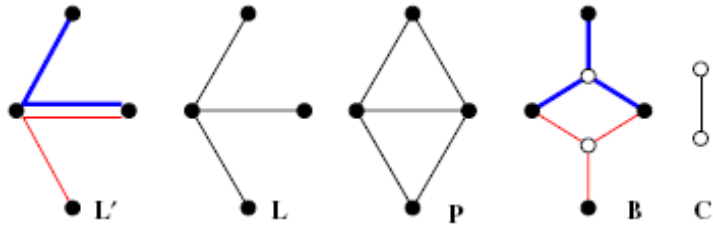

**Fig. 1.** Representation of a PTN in graph form. **L'**-space: filled circles represent stations serviced by two different routes shown by a bold and a thin line. **L**-space: reduction of **L'** to a simple graph with single links. **P**-space: any two stations that are serviced by a common route are linked. **B**-space: stations are linked to the routes (open circles) they belong to. **C**-space: routes are connected when they share a common station

To generate this knowledge we performed a thorough analysis of the behaviour under attack of PTNs of 14 major cities of the world, taking cities from different continents, with different concepts of planning and different history. In our sample, the number of stations $N$ and routes $M$ ranged from $N=1544$, $M=124$ (Düsseldorf) to $N = 46244$, $M = 1893$ (Los Angeles). This allowed us to produce a reliable statistics and to access some unique features of the networks under consideration.

There are many different ways to represent a PTN of a city in a form of a graph, some of such representations – 'spaces' - are shown in Figs. 1, 2. While the primary network topology is defined by a set of routes each servicing an ordered series of given stations (**L'**-space, see Fig.1, left), a number of different neighbourhood relations may be defined both for the routes and the stations. E.g. one either defines two stations as neighbours whenever they are serviced by a common route (**P**-space) or one considers networks formed by routes (nodes) that are connected if they share a common station (**C**-space). In our analysis we made use of all the above representations, first looking on different characteristics of PTN calculated for the corresponding graphs and then performing attacks and measuring changes in these characteristics. The implementation (and effect) of an attack differs in different spaces. To give two examples: attacks in the **L**-space correspond to situations, in which given public transport stations and all their incident links cease to operate for all means of traffic that go through them, whereas if a station-node is removed in the **P**-space, the corresponding stop of the route is cancelled while the route otherwise keeps operating. The above described attacks refer to the removal of nodes. Alternatively, we have analyzed the behaviour of PTN when the links are removed [14].

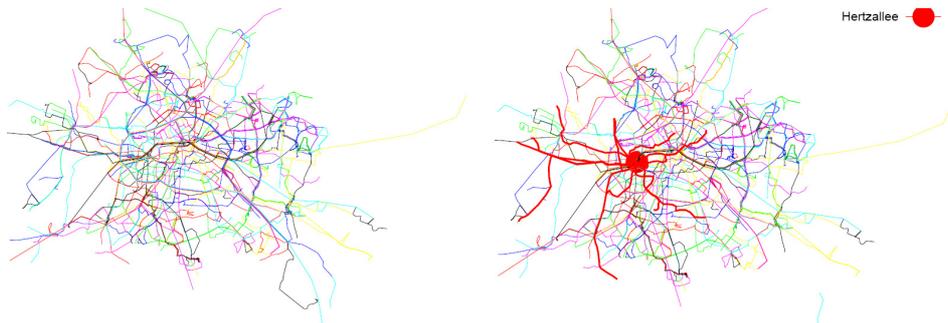

**Fig. 2.** The PTN of Berlin (left): Interpreting the stations as vertices and the lines as links (identifying multiple edges) results in an **L**-space graph. Right: The sub-network of nodes (links shown in bold red) that can be reached from Hertzallee (shown as a bold red spot) without changing mode of transport corresponds to the **P**-space neighbourhood of that station.

With given PTN representation in the form of a graph we are in the position to describe the observables we will be interested in as well as to describe the way attacks of different strategies are performed.

## 3. Observables and attack strategies

In practice, the origin of the attack and its scenario may differ to large extent, ranging from random failure, when a node or a link in a network is removed at random to a targeted destruction, when the most influential network constituents are removed according to their operating characteristics. Moreover, choosing a certain criterion, one can prepare the list of the nodes for the *initial* network and remove the nodes according to this list. Alternatively, one can continuously measure nodes (or links) characteristics after each step and modify such a list in the course of an attack. Attacks according to recalculated lists often turn out to be more harmful than the attack strategies based on the initial list, suggesting that the network structure changes as important nodes or links are removed [16,17].

One can single out two different impacts to the effectiveness of a network and its resilience during an attack. The first one has purely topological origin and is uniquely defined by network structure, the second one originates from the load on a network, i.e. it takes into account intensity of the transportation processes. In the particular case of a PTN this second impact is characterised by the number of vehicles or number of passengers that use a given route. In our study we will be interested only in the first impact, primary addressing the network topology and leaving aside its load. That is, speaking about network robustness to an attack we will first of all mean how 'complete' remains a PTN when its constituents are removed. There are different observables that are usually employed to characterise such robustness. In particular, these are the mean shortest path length $\langle l \rangle$, mean of its inverse $\langle l^{-1} \rangle$, size of the largest component $S$ [12,13,16]. Below, we will exploit the last quantity defined as

$$S = N_1 / N, \qquad (1)$$

where $N$ and $N_1$ are numbers of nodes of the network and of its largest component correspondingly. In practice, for a finite network, such a quantity serves as an analogue of a giant connected component (GCC) which is defined for an infinite network only [1]. In turn, the GCC serves as the analogue of a percolation cluster, when the problem of network resilience is treated in terms of percolation theory [18].

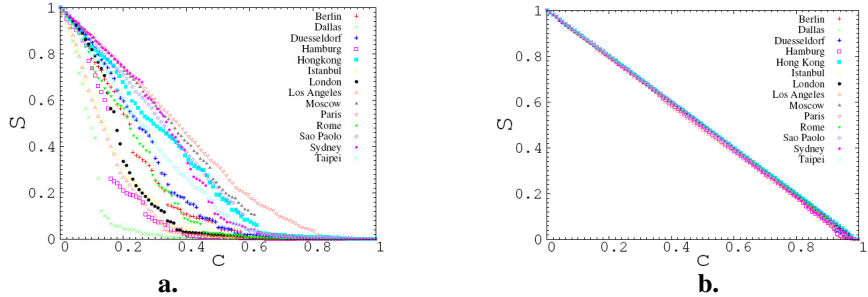

**Fig. 3.** The largest cluster size $S$ of different PTN as function of the fraction of removed nodes $c$. Random scenario. **a: L**-space, **b: P**-space

In Fig. 3 we show how $S$ changes as function of the concentration of removed nodes $c$ when these nodes are removed randomly without any reference to their characteristics. Below, we will call such scenario a random one and denote it as RV (random vertex). The data is displayed for the PTNs of 14 different cities, as listed in the corresponding legends. Fig. 3**a** shows results for the PTN represented as a graph in the **L**-space, whereas Fig. 3**b** shows data for the **P**-space. One immediately notes that the reaction of the **P**-space graphs on random attacks is rather homogeneous and merely corresponds to continuous linear decrease of $S(c)$. This is easy to understand if one recalls that in the **P**-space each route enters the PTN representation as a complete graph and hence a random removal of any station node does not cause network segmentation. On the contrary, the reaction of the PTN graphs on a random attack in the **L**-space ranges from abrupt breakdown (Dallas) to a slow almost linear decrease (Paris).

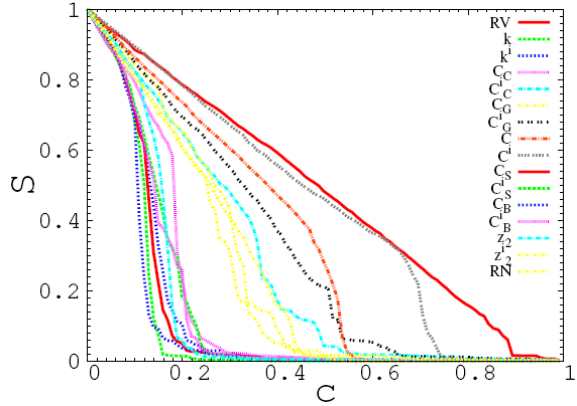

**Fig. 4.** Largest component size $S$ of the PTN of Paris as function of the fraction of removed nodes $c$ for different attack scenarios. Each curve corresponds to a different scenario as indicated in the legend. Lists of removed nodes were prepared according to their degree $k$, closeness $C_C$, graph $C_G$, stress $C_S$, and betweenness $C_B$ centralities, clustering coefficient $C$, and next nearest neighbors number $z_2$. A superscript $i$ refers to lists prepared for the initial PTN before the attack. RV and RN denote the removal of a random vertex or of its randomly chosen neighbour, respectively

Before discussing correlations between the characteristics of unperturbed PTN and their robustness to attacks let us first explain different attack strategies we pursued in our study. In the network literature different attack scenarios are used to analyse the resilience of a complex network [12-14,16,17]. Generally, these are based on the intuitive assumption that the largest impact on a network is caused by the removal of its most important constituents. To quantify such importance of a node, one often uses the node degree $k$ (i.e. the number of nearest neighbours of a given node), closeness $C_C$, graph $C_G$, stress $C_S$, and betweenness $C_B$ centralities (see e.g. [19] for definitions and discussion ). To give an example, for a given node $j$ the last quantity is defined as

$$C_B(j) = \sum_{s \neq j \neq t \in N} \frac{\sigma_{st}(j)}{\sigma_{st}}, \qquad (2)$$

where $\sigma_{st}(j)$ is the number of shortest paths between nodes $s$ and $t$ that belong to the network $N$ and go through the node $j$. One can also measure the importance of a given node by the number of its second nearest neighbours $z_2$ or its clustering coefficient $C$. The latter is the ratio of the actual number of links between the node's nearest neighbours and the maximal possible number of mutual links between them. In our analysis we made use of different attack scenarios, removing the nodes according to the lists ordered by decreasing node degree $k$, centralities $C_C$, $C_G$, $C_S$, $C_B$, number of second nearest neighbours $z_2$, and increasing clustering coefficient $C$. Such lists were either prepared for an unperturbed network or recalculated after each step of attack. Together, this makes fourteen different attack scenarios which we completed by the above described random attack (RV) as well as by a scenario where a randomly chosen neighbour of a randomly chosen node (RN) is removed. The last scenario appears to be effective for immunization problems [20].

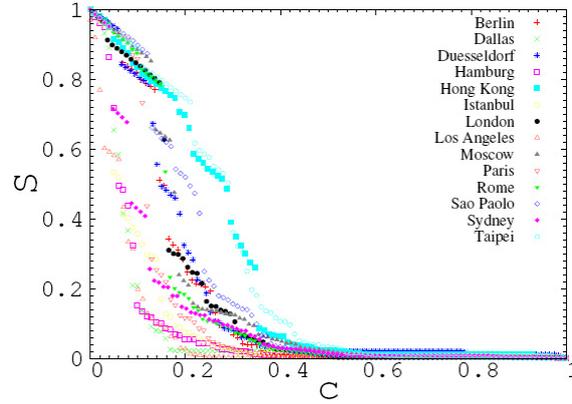

**Fig. 5.** Size of the largest cluster **S** as a function of the fraction of removed nodes **c**. **P**-space, highest betweenness scenario (recalculated)

A typical result of our study is shown in Fig. 4. There, we show changes in the largest component size $S$ of the PTN of Paris as a function of the removed nodes fraction $c$ for the above described attack scenarios. Each curve corresponds to a different scenario as indicated in the legend. A similar analysis was performed for all other PTN from our database. As expected, it appears that the impact of an attack for a given PTN graph crucially depends on the attack scenario. Moreover, the most harmful scenarios differ for different graph representations (different 'spaces'). In particular, for the **L**-space graphs the most harmful scenarios are those defined by the node degree, betweenness, closeness and stress centralities, and second nearest neighbours number whereas for the **P**-space graphs the node degree does not play such an important role and the most destructive are centrality-oriented scenarios. In Fig. 5 we show the size of the largest cluster $S$ of different PTN graphs in **P**-space as a function of the fraction of removed nodes $c$ for the recalculated highest betweenness scenario which appears to be the most destructive for the **P**-space graphs. Indeed, the special role played by the highest betweenness nodes is explained by the fact, that they join different routes (represented in **P**-space by separate complete graphs) and their removal leads to rapid network segmentation.

Another instance of our studies is given by Fig. 6. There we show the behaviour of the largest cluster $S$ of different PTN observed when links are removed at random between station nodes in **L**-space. For each of the PTN randomly removing nodes appears to be more effective as randomly removing links [14].

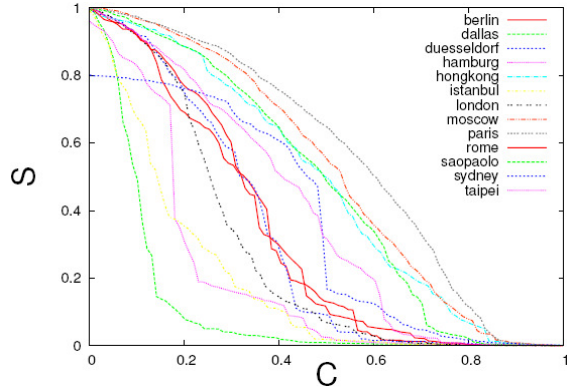

**Fig. 6.** Size of the largest cluster S as a function of the fraction of removed links c. **L**-space, random scenario

## 4. Conclusions and outlook

In our analysis we made an attempt to identify correlations between characteristics of the unperturbed PTN and their robustness to attacks. In particular, we have found [10], that some of the PTN under consideration manifest scale-free properties: their node degree distribution is governed by a power-law decay

$$P(k) \sim k^{-\gamma}. \qquad (3)$$

In the case of the **L**-space, the scale-free behaviour was markedly observed for 7 PTN out of 14, in the **P**-space these were 4 out of 14. An exponential decay of $P(k)$ observed for the other PTN, nevertheless may be fitted by a power law (3) with less accuracy. Our analysis verifies correlations between the numerical value of the node-degree distribution exponent $\gamma$ and the segmentation concentration of PTN (the last was numerically estimated on the base of different methods in [12,13], see also Fig. 7 for an illustration). In particular, both for the RV and the recalculated node-degree attack scenarios in **L**-space we have found that a small value of $\gamma$ corresponds to higher PTN robustness. Notable examples are given by PTN of Paris ($\gamma = 2.62$), Saõ Paolo ($\gamma = 2.72$), and Hong Kong ($\gamma = 2.99$). The segmentation concentration for these PTN at the RV scenario it is $c_s = 0.38; 0.32; 0.30$, respectively. Alternatively, at the same attack scenario PTNs of London ($\gamma = 4.48$), Los Angeles ($\gamma = 4.85$), and Dallas ($\gamma = 5.49$) have much lower segmentation concentration: $c_s = 0.175; 0.130; 0.090$.

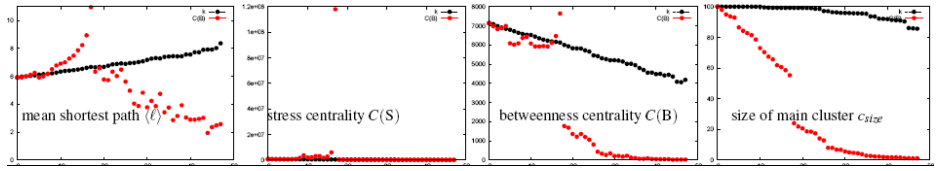

**Fig. 7.** Search of the segmentation concentration during an attack scenario. From left to right: changes in the mean shortest path length, stress and betweenness centralities and size of the main cluster as functions of $c$ for the highest node degree (black curves) and highest betweenness centrality scenarios for the Berlin PTN.

Another instructive observation concerns the applicability of the Molloy-Reed criterion [21] which has been formulated for networks with given node degree distribution but otherwise random linking between vertices. For such equilibrium networks a GCC was shown to be present if

$$\langle k(k-2) \rangle \geq 0, \tag{4}$$

where angular brackets stand for a network average in the limit of an infinite network with given $P(k)$. Therefore, a GCC is absent for small values of the parameter $\kappa \equiv \langle k^2 \rangle / \langle k \rangle < 2$. Calculating values of $\kappa$ for the unperturbed PTN we have found that for these real-world networks smaller values of $\kappa$ in general indicate a smaller segmentation concentration $c_s$ both for the RV and for the recalculated node-degree attack scenarios in **L**-space. Alternatively, the mean shortest path length appears to be a useful indicator for PTN robustness to attacks in **P**-space. As we have shown [13], high resilience of a PTN in **P**-space in the highest betweenness centrality attack scenario is indicated by a small shortest path length of the unperturbed network.

## Acknowledgements

The paper is based on an invited lecture given by one of us (Yu.H.) at the Workshop „NET 2009: evolution and complexity" (Rome, May 28th-30th, 2009). The hospitality of the organizers is kindly acknowledged. This work was supported by the following institutions and projects: Austrian Fonds zur Förderung der wissenschaftlichen Forschung under Project No. P19583-N20 (Yu.H.), cooperation programme 'Dnipro' between the Ministry of Foreign Affairs of France and the Ministry of Education and Science of Ukraine (B.B. & Yu.H); Ecole Doctorale EMMA of the University Nancy and an Overseas Research Student Scholarship (T.H.), and the Applied Research Fellow Scheme of Coventry University (CvF).